\documentclass{nature}

\usepackage{amsmath}
\usepackage[pdftex]{graphicx}
\usepackage{dcolumn}  
\usepackage{bm}           
\usepackage{dsfont}
\usepackage{color}
\usepackage{braket}
\usepackage{booktabs}
\usepackage{hyperref}

\newcounter{defcounter}
\usepackage{amsthm}

\newcommand{\Rmnum}[1]{\uppercase\expandafter{\romannumeral #1}}

\title{Long-lived Coherent Phonons Reveal Competing Thermal and Many-Body Dynamics in $\alpha$-In$_{2}$Se$_{3}$} 

\author{Xuanchao Zhang$^{1,*}$, Dehao Yuan$^{1,*}$, Junhua Zhou$^{1}$, Vandana Tiwari$^{2}$, Fulu Zheng$^{1}$, Ajay Jha$^{3,4}$, Hong-Guang Duan$^{1}$} 

\begin{document} 

\maketitle 

\begin{affiliations} 
\item Department of Physics, School of Physical Science and Technology, Ningbo University, Ningbo, 315211, P.R. China 
\item Linac Coherent Light Source, SLAC National Accelerator Laboratory, Menlo Park, California 94025
\item Rosalind Franklin Institute, Harwell, Oxfordshire OX11 0QX, United Kingdom 
\item Department of Pharmacology, University of Oxford, Oxford, OX1 3QT United Kingdom  \\ 
\centerline{\underline{\date{\bf \today}}} 
\end{affiliations} 

\begin{abstract} 

The fate of quantum coherence following photoexcitation is a central problem in nonequilibrium condensed-matter physics, especially in low-dimensional solids where electronic, structural and many-body energy scales are strongly coupled. Here, we investigate this interplay in ferroelectric $\alpha$-In$_{2}$Se$_{3}$ using broadband transient-grating spectroscopy with $\sim$5 fs laser pulses. Photoexcitation launches pronounced oscillations dominated by a $\sim$104 cm$^{-1}$ mode that persists for several picoseconds. Measurements from 10 to 300 K show that its frequency remains nearly unchanged while its coherence is progressively suppressed, distinguishing the lattice coordinate from thermally activated dephasing. Varying excitation energy reveals a distinct crossover in which increasing photoexcitation progressively modifies and damps the coherent response. First-principles calculations assign the dominant oscillation to a 101.47 cm$^{-1}$ $\Gamma$-point optical phonon involving collective In-Se displacement, while quantum-dynamical simulations reproduce the main transient-grating oscillations. These results establish $\alpha$-In$_{2}$Se$_{3}$ as a model system for independently probing lattice coherence, thermal fluctuations and carrier-density-dependent interactions, revealing how a polar van der Waals semiconductor evolves from coherent lattice motion toward an incoherent many-body state after ultrafast excitation. 

\end{abstract}


Understanding how a quantum solid evolves after photoexcitation is a central problem in condensed-matter physics. An ultrashort pulse creates a nonequilibrium electronic distribution that exchanges energy and momentum through carrier-carrier, carrier-phonon and phonon-phonon interactions.\cite{Ref1} These processes determine whether the excitation rapidly becomes incoherent or launches collective lattice motion that retains phase information. \cite{Ref1, Ref2, Ref3, Ref4, Ref5, Ref6} Coherent phonons therefore reveal how electronic excitation couples to specific structural coordinates and how this coherence is destroyed by the many-body environment.\cite{Ref2,Ref7,Ref8} This question is especially important in low-dimensional and polar materials, where reduced screening, structural anisotropy and strong electron-lattice coupling reshape relaxation pathways.

Layered van der Waals materials provide a rich setting for studying these effects.\cite{Ref9,Ref10,Ref11,Ref12} Their weak interlayer coupling and strong intralayer bonding produce anisotropic electronic structures and low-energy lattice modes that are sensitive to temperature, carrier density, strain and stacking. In$_{2}$Se$_{3}$ is particularly suitable because it supports multiple polymorphs and stacking configurations, including $\alpha$, $\beta$/$\beta'$ and $\gamma$ phases with distinct symmetry and coordination.\cite{Ref13,Ref14,Ref15,Ref16,Ref17} $\alpha$-In$_{2}$Se$_{3}$ consists of covalently bonded Se--In--Se--In--Se quintuple layers separated by van der Waals gaps. Its broken inversion symmetry produces coupled in-plane and out-of-plane polarization, making it a prototypical two-dimensional ferroelectric semiconductor.\cite{Ref18,Ref19,Ref20,Ref21} Ferroelectricity persists to the few-layer limit, while stacking, strain, domains and electric fields modify its polarization and electronic properties.\cite{Ref22,Ref23,Ref24} Thus, its lattice is not merely a thermal reservoir: atomic displacement is directly linked to symmetry, polarization and electronic structure.

This sensitivity is reflected in the vibrational spectrum. Raman measurements identify low-frequency modes near $\sim$90-110 cm$^{-1}$ and higher-frequency modes around $\sim$180-200 cm$^{-1}$, whose frequencies and intensities fingerprint the $\alpha$ polymorph and stacking configuration.\cite{Ref15,Ref25,Ref26} The $\sim$100 cm$^{-1}$ region is especially relevant because it contains collective In--Se motion associated with the polar lattice. Structural, pressure-dependent and theoretical studies further indicate strong coupling between lattice coordinates and the energetic landscape of In$_{2}$Se$_{3}$.\cite{Ref15,Ref16} Coherent excitation of these modes could therefore directly probe communication between the electronic system and the symmetry-breaking lattice coordinate.

Ultrafast studies have focused mainly on carrier populations rather than coherent lattice dynamics. Transient-absorption and transient-reflectivity measurements have identified band filling, band-gap renormalization, carrier cooling, trapping and density-dependent Auger recombination in $\alpha$-In$_{2}$Se$_{3}$.\cite{Ref27,Ref28,Ref29} Increasing excitation density accelerates carrier relaxation and enhances Auger contributions\cite{Ref27,Ref28}, while recent measurements report unusually slow hot-carrier cooling over tens of picoseconds\cite{Ref29}. These results establish a strongly interacting nonequilibrium carrier population, but the coherent structural response remains less understood: which lattice coordinates are launched, how long their phase memory survives, and whether thermal phonons and photoinduced carrier interactions destroy coherence through the same pathway. 

Here we address this question by independently varying temperature and pump-fluence while resolving the coherent response of $\alpha$-In$_{2}$Se$_{3}$. Temperature controls phonon occupation and anharmonic scattering, whereas pump-fluence changes the photocarrier population and carrier-mediated interactions. Their combined variation distinguishes lattice-controlled dephasing from density-driven electronic scattering and allows us to ask: how does a photoexcited polar semiconductor evolve from phase-correlated carrier-lattice motion to an incoherent many-body regime, and which lattice coordinate mediates this transition? 

This question is investigated using broadband transient-grating spectroscopy\cite{Ref30,Ref31} with $\sim$5 fs excitation pulses to follow the electronic and vibrational response of $\alpha$-In$_{2}$Se$_{3}$ with 3R structure from the earliest photoexcited state to picosecond lattice relaxation. Temperature is varied between 10 and 300 K to control the thermal phonon environment, while pump-fluence is independently varied to tune the photocarrier density. Fourier and continuous-wavelet analyses resolve the frequencies and temporal envelopes of the resulting coherent oscillations. Density-functional lattice-dynamics calculations identify the atomic motions associated with the experimentally observed modes, and an effective vibronic open-quantum-system model treated using hierarchical equations of motion connects these lattice coordinates with the measured transient-grating response.

\section*{Results} 

\subsection{Coherent lattice response.}

\begin{figure}[h!]
\begin{center}
\includegraphics[width=15.0cm]{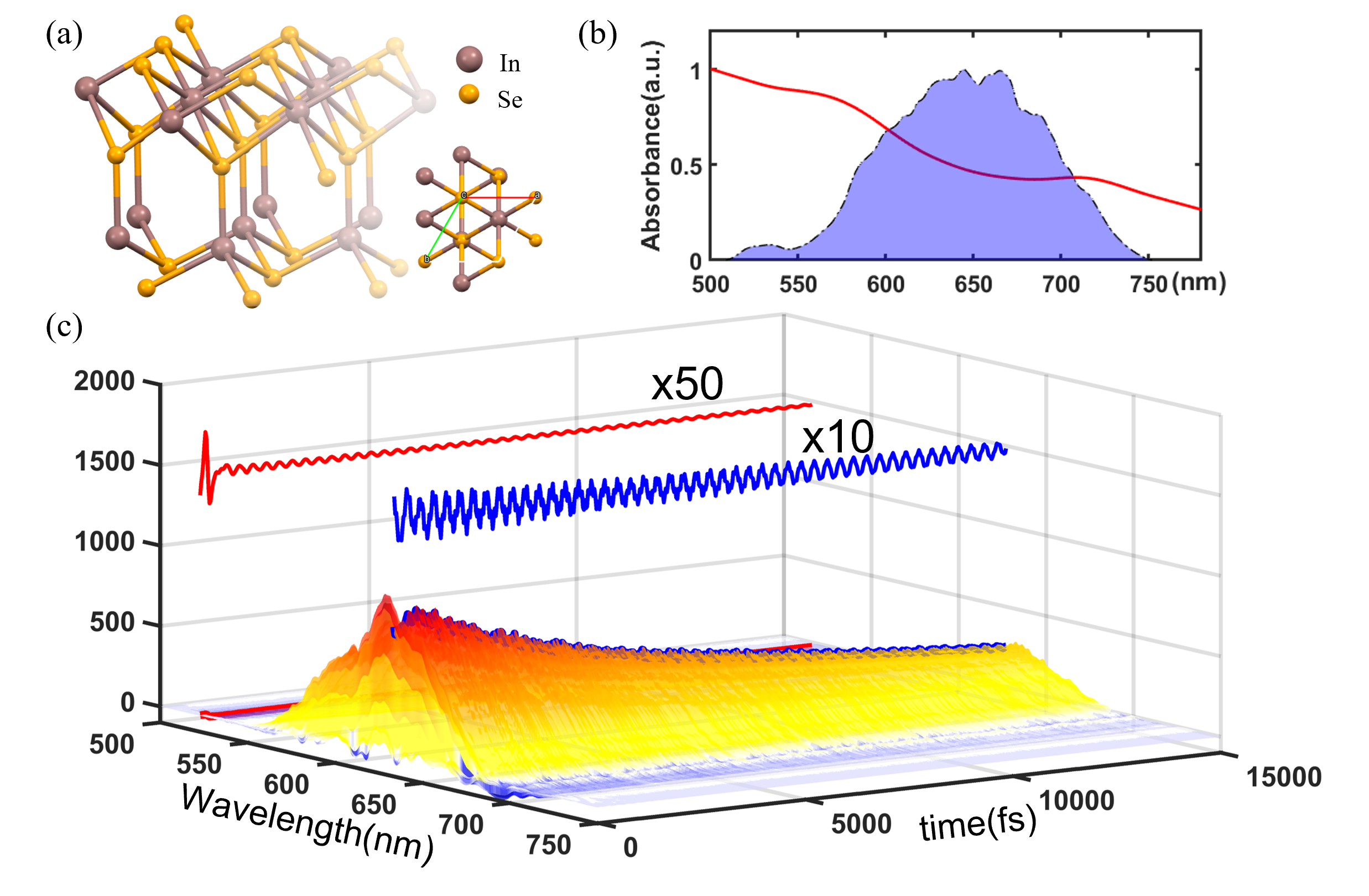} 
\caption{\label{fig:Fig1} \textbf{Ultrafast excitation reveals coherent lattice dynamics in $\alpha$-In$_2$Se$_3$.} (a) Crystal structure of layered $\alpha$-In$_2$Se$_3$ showing the Se-In-Se-In-Se quintuple-layer motif. (b) Steady-state absorption spectrum together with the broadband excitation spectrum used for transient-grating measurements. (c) Room-temperature transient-grating response spanning 500-750 nm. Representative traces extracted from distinct spectral regions reveal pronounced oscillatory modulations superimposed on the electronic population dynamics, demonstrating coherent nuclear motion persisting for multiple vibrational periods after photoexcitation.} 
\end{center}
\end{figure}

$\alpha$-In$_{2}$Se$_{3}$ is a polar van der Waals semiconductor in which electronic excitation occurs within asymmetric Se-In-Se-In-Se quintuple layers. Fig.\ \ref{fig:Fig1}(a) shows the layered structure, while Fig.\ \ref{fig:Fig1}(b) compares the steady-state absorption spectrum with the broadband excitation window. The $\sim$200 nm pulse bandwidth accesses a broad range of electronic transitions while providing the temporal resolution needed to resolve low-frequency lattice motion. 

The 300 K transient-grating spectrum is shown in Fig.\ \ref{fig:Fig1}(c). Although the response extends across approximately 500-750 nm, it cannot be described by monotonic population relaxation alone. Pronounced oscillations persist for several vibrational cycles and appear in both positive and negative transient features, with wavelength-dependent amplitudes and lifetimes. The response therefore contains a slowly varying electronic population and an oscillatory contribution from a periodically evolving nuclear coordinate. Because the excitation pulse is much shorter than the dominant phonon period, this lattice motion is launched impulsively and can be followed directly in time.

\begin{figure}[h!]
\begin{center}
\includegraphics[width=17.0cm]{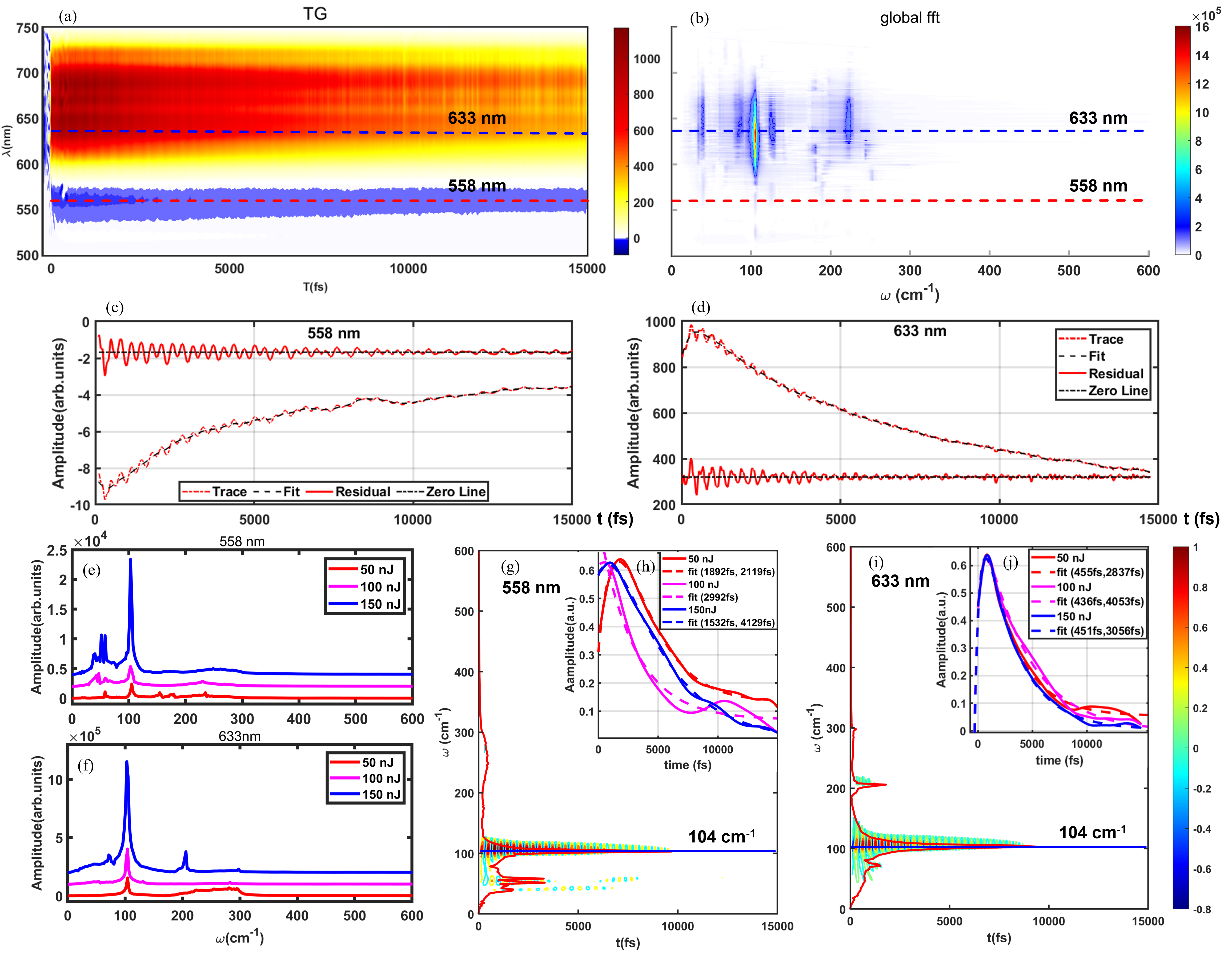} 
\caption{\label{fig:Fig2} \textbf{Excitation-density dependence of the coherent phonon response.} (a) Room-temperature transient-grating spectrum highlighting the spectral regions near 558 and 633 nm. (b) Global Fourier analysis revealing a dominant $\sim$104 cm$^{-1}$ oscillation. (c,d) Population kinetics, fitted backgrounds and oscillatory residuals. (e,f) Fourier spectra at different pump-fluence. (g,i) Wavelet maps resolving the temporal evolution of the phonon wavepacket. (h,j) Extracted buildup and decay times demonstrate wavelength-dependent sensitivity of lattice coherence to photocarrier density.} 
\end{center}
\end{figure}

Figure 2 shows the frequency content and pump-fluence dependence. Distinct positive and negative features occur near 633 and 558 nm (Fig.\ \ref{fig:Fig2}a). Fourier analysis reveals a narrow contribution near 104 cm$^{-1}$ across the transient spectrum (Fig.\ \ref{fig:Fig2}b). After removing the non-oscillatory population dynamics (Fig.\ \ref{fig:Fig2}c, d), this mode dominates the residual response, with weaker components below 100 cm$^{-1}$ and near 180-200 cm$^{-1}$. Wavelet analysis further shows that the 104 cm$^{-1}$ feature is a time-dependent wavepacket rather than a static Raman contribution (Fig.\ \ref{fig:Fig2}g, i). Its amplitude rises after excitation and decays over several picoseconds. The same frequency in different spectral regions indicates that a common lattice coordinate modulates multiple electronic transitions. 

Increasing pump fluence changes both the buildup and damping of the coherent oscillation. At the shorter-wavelength feature, the buildup time increases from $\sim$257 fs at 50 nJ (0.38 mJ cm$^{-2}$) to $\sim$537 fs at 100 nJ (0.75 mJ cm$^{-2}$) and $\sim$1.2 ps at 150 nJ (1.13 mJ cm$^{-2}$), while the coherent envelope becomes more strongly damped at the highest pump fluence. In contrast, the response at 633 nm exhibits a weaker dependence on pump fluence. Thus, increasing photoexcitation does not simply scale the amplitude of a fixed phonon response but instead modifies both the buildup and damping of the coherent lattice dynamics, indicating a pump-fluence-dependent interaction between the photoexcited electronic population and coherent lattice motion. Previous ultrafast measurements of $\alpha$-In$_{2}$Se$_{3}$ have shown that increasing photocarrier density enhances carrier-carrier and Auger-type relaxation channels \cite{Ref27,Ref28}. The present observations are therefore consistent with increasing competition between coherent lattice motion and density-dependent many-body relaxation, although the TG measurements alone do not uniquely identify the microscopic scattering channel responsible for the additional damping.

\subsection{Thermal dephasing.} 

\begin{figure}[h!]
\begin{center}
\includegraphics[width=12.0cm]{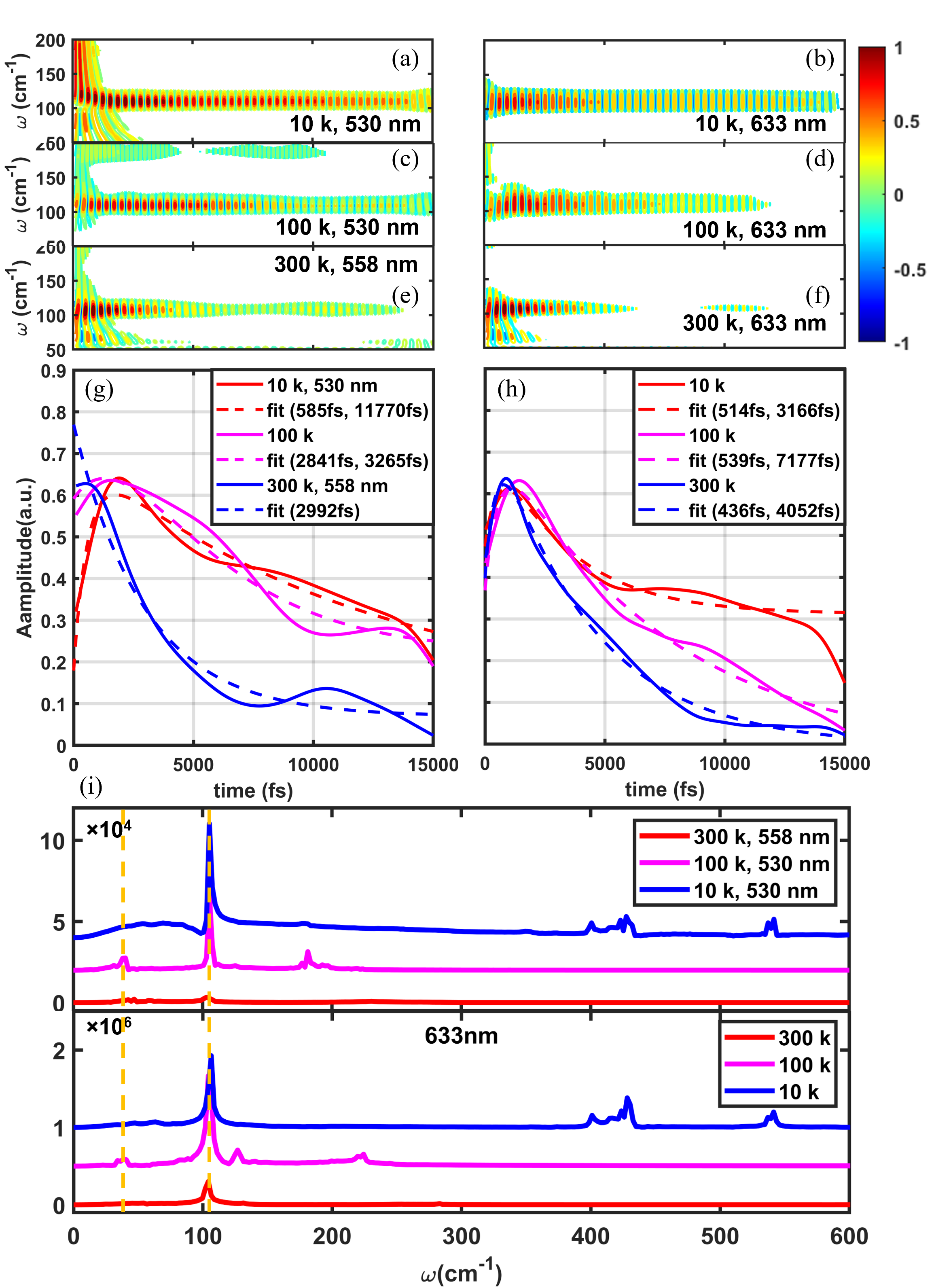} 
\caption{\label{fig:Fig3} \textbf{Thermal control of phonon phase memory in $\alpha$-In$_2$Se$_3$.} Time-frequency wavelet maps of the oscillatory transient-grating response measured between 10 and 300 K at representative probe wavelengths, together with extracted temporal envelopes and Fourier spectra. The dominant mode remains centered near $\sim$104 cm$^{-1}$ across the complete temperature range, whereas its temporal persistence increases strongly on cooling. The contrasting temperature dependences of frequency and damping identify thermal dephasing, rather than a change in vibrational identity, as the principal effect of the phonon environment.} 
\end{center}
\end{figure}

If the oscillations arise from genuine lattice coherence, temperature should affect their lifetime more strongly than their frequency. Fig.\ \ref{fig:Fig3} tests this expectation by comparing the response from 10 to 300 K. At all temperatures, the oscillations remain centred near 104 cm$^{-1}$, showing that cooling does not select a different vibrational mode. Instead, temperature controls how long the same collective coordinate retains phase memory. At 10 K, the wavepacket persists substantially longer, whereas increasing temperature shortens its lifetime and broadens the Fourier response. The distinction between frequency and damping is significant. The frequency reflects the curvature of the lattice potential, while the damping reflects interactions with surrounding degrees of freedom. The weak temperature dependence of the frequency and strong dependence of the lifetime therefore identify dephasing, rather than a change in phonon identity, as the dominant thermal effect.

Wavelet maps show an extended $\sim$104 cm$^{-1}$ ridge at cryogenic temperature, earlier amplitude loss at 100 K and stronger temporal confinement at 300 K. Fourier spectra show a corresponding enhancement and sharpening of the coherent contribution upon cooling. This behaviour is consistent with thermally activated dephasing, for which anharmonic phonon-phonon scattering provides a natural microscopic contribution. Lower thermal phonon occupation decreases the phase space for phonon-phonon interactions, allowing the optical mode to retain coherence for more oscillation periods. Heating restores these scattering channels and accelerates phase randomisation without substantially changing the harmonic frequency. Temperature and pump-fluence therefore provide independent controls. Density modifies the interacting electronic population, whereas temperature changes the thermal lattice environment. Because both perturbations shorten the coherent response through different variables, the 104 cm$^{-1}$ wavepacket lies at the interface between electronic and lattice many-body dynamics.

\subsection{Microscopic phonon assignment.} 

We identify the microscopic coordinate responsible for the oscillations using density-functional lattice-dynamics calculations. Several $\Gamma$-point optical modes occur in the relevant frequency range, including calculated modes at 101.47, 152.37, 179.82 and 236.98 cm$^{-1}$, with distinct in-plane and out-of-plane In and Se displacements. The best agreement with experiment is obtained for the 101.47 cm$^{-1}$ mode, only a few wavenumbers below the observed 104 cm$^{-1}$ oscillation. Its eigenvector contains substantial collective displacement of the In-Se framework normal to the layers, consistent with Raman-active modes reported near $\sim$104-110 cm$^{-1}$ in $\alpha$-In$_{2}$Se$_{3}$.\cite{Ref15,Ref25} This assignment is especially relevant for a polar layered semiconductor. Because the electronic structure and ferroelectric polarization are sensitive to relative In-Se displacement, coherent motion along this coordinate can periodically modulate electronic transition energies and oscillator strengths. The same phonon appearing in multiple spectral regions therefore reflects a common structural coordinate coupled to several optical transitions.

\begin{figure}[h!]
\begin{center}
\includegraphics[width=12.0cm]{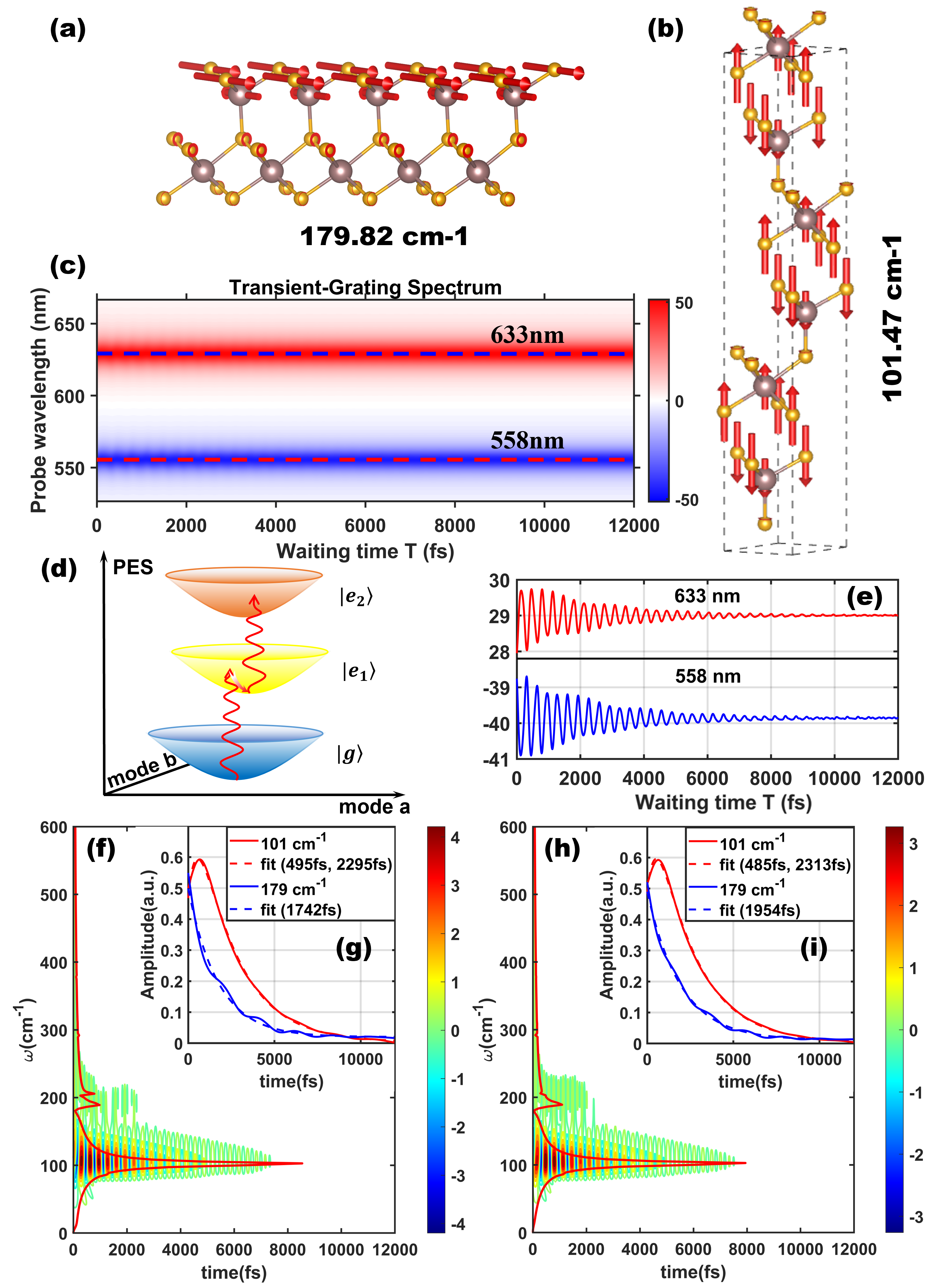} 
\caption{\label{fig:Fig4} \textbf{Microscopic assignment and quantum-dynamical description of the coherent lattice coordinate.} (a-b) Calculated $\Gamma$-point phonon eigenvectors of $\alpha$-In$_2$Se$_3$ at 101.47 and 179.82 cm$^{-1}$. (c) Simulated transient-grating spectrum. (d) Schematic potential-energy surfaces of the three-level, two-mode model. (e) Calculated waiting-time traces reproduce coherent modulations in the 558 and 633 nm spectral regions. (f,h) Population dynamic maps of the baseline TG waiting-time traces at 558~nm and 633~nm, showing components near 104 and 179~cm$^{-1}$. (g,i) Corresponding wavelet amplitudes of the 104~cm$^{-1}$ (red) and 179~cm$^{-1}$ (blue) components. Solid curves show the extracted amplitudes and dashed curves show empirical fits. } 
\end{center}
\end{figure}

We next test the dynamical consistency of this assignment using an effective three-level, two-mode vibronic model. The displaced potential-energy surfaces of the electronic ground state $\ket{g}$ and excited manifolds $\ket{e_1}$ and $\ket{e_2}$ are illustrated in Fig.~\ref{fig:Fig4}(d). The baseline Hamiltonian employs nominal vibrational frequencies of 101.47 and 187.7~cm$^{-1}$ to represent the low- and high-frequency coordinates, respectively; for the finite-duration calculated traces, the corresponding wavelet maxima occur near 101 and 179~cm$^{-1}$. Environmental relaxation and dephasing are treated nonperturbatively using hierarchical equations of motion. The calculated transient-grating spectrum contains oppositely signed features near 633 and 558~nm in Fig.~\ref{fig:Fig4}(c), and the waiting-time traces extracted at these wavelengths show damped coherent modulations in Fig.~\ref{fig:Fig4}(e). Their continuous-wavelet-transform maps at 558 and 633~nm in Fig.~\ref{fig:Fig4}(f,h), respectively, resolve both the $\sim101$ and $\sim179$~cm$^{-1}$ components. The corresponding amplitude profiles and empirical fits are shown in Fig.~\ref{fig:Fig4}(g,i). At 558~nm, the fitted constants are 495 and 2295~fs for the $\sim101$~cm$^{-1}$ component and 1742~fs for the $\sim179$~cm$^{-1}$ component; at 633~nm, the corresponding values are 485 and 2313~fs, and 1954~fs. Their similarity at the two wavelengths is consistent with both spectral features being modulated by the same model coordinates. These quantities characterize the simulated wavelet envelopes and should not be interpreted as independently measured or unique microscopic phonon lifetimes. Because the transition energies and vibrational frequencies are inputs to this phenomenological model, the calculation establishes dynamical consistency and shows that the selected coordinate set is sufficient to generate the principal response, rather than independently proving the assignment or excluding contributions from other modes. The full Hamiltonian, wavelet procedure, fitting functions, and mode-substitution test are provided in Sec.~\Rmnum{7} of the SI.


\section*{Discussion} 

The combined temperature-, density- and frequency-resolved measurements reveal nonequilibrium interactions that are obscured when only carrier populations are considered. Previous studies identified band filling, band-gap renormalization, hot-carrier cooling and Auger recombination as important relaxation channels in $\alpha$-In$_{2}$Se$_{3}$.\cite{Ref27, Ref28, Ref29, Ref32} Our results add a structural dimension: throughout relaxation, the electronic excitation remains coupled to a well-defined collective lattice coordinate near 104 cm$^{-1}$ for several vibrational periods. The near-invariant frequency between 10 and 300 K, together with strong changes in temporal persistence, separates the lattice eigenmode from the bath responsible for dephasing. This behaviour indicates that anharmonic interactions affect coherence more strongly than they renormalize the harmonic potential.\cite{Ref33,Ref34,Ref35} Cooling reduces thermal phonon occupation and accessible phonon-phonon scattering channels, allowing the coherent wavepacket to persist. Temperature therefore provides a direct measure of the lattice contribution to decoherence. 

Pump-fluence probes a complementary many-body regime. Earlier work showed that increasing carrier density enhances Auger recombination and alters the balance between band filling and band-gap renormalization.\cite{Ref27,Ref28} Here, it also modifies the formation and damping of the coherent phonon response, demonstrating that lattice dynamics remain coupled to the electronic population. $\alpha$-In$_{2}$Se$_{3}$ therefore passes through an intermediate regime in which electronic and lattice degrees of freedom remain dynamically correlated before scattering produces an incoherent quasithermal state.

Assigning the dominant oscillation to the calculated 101.47 cm$^{-1}$ In-Se mode is significant because structural displacement in $\alpha$-In$_{2}$Se$_{3}$ is linked to broken inversion symmetry and ferroelectric polarization. We do not infer ultrafast polarization switching. Instead, photoexcitation coherently drives a structural coordinate within the low-energy lattice manifold governing the material’s polar properties. Selected phonon modes may therefore provide transient structural pathways for controlling electronic and ferroic behaviour, rather than acting solely as sources of decoherence. Thus, $\alpha$-In$_{2}$Se$_{3}$ offers a useful platform for distinguishing thermal and carrier-density-driven routes from coherent dynamics to statistical relaxation. More broadly, it connects ultrafast spectroscopy with the question of how collective coordinates emerge, persist and decohere in an interacting quantum solid.


\section*{Conclusion} 

Our measurements reveal how electronic excitation reorganizes the lattice of $\alpha$-In$_{2}$Se$_{3}$ in real time. Broadband transient-grating spectroscopy resolves a dominant $\sim$104 cm$^{-1}$ optical phonon, which first-principles calculations assign to a collective In-Se displacement. An effective vibronic quantum-dynamical model reproduces the principal oscillatory characteristics, demonstrating the dynamical consistency of this assignment. These results identify an intermediate regime of carrier relaxation in which lattice motion retains phase coherence for several vibrational periods. Temperature controls the lifetime of this coherence through the thermal phonon environment, whereas pump-fluence modifies its competition with carrier-mediated many-body scattering. By independently varying these parameters, we distinguish thermal dephasing from density-dependent electronic relaxation, processes that are often intertwined in conventional room-temperature pump-probe measurements. Our findings establish $\alpha$-In$_{2}$Se$_{3}$ as a model system for studying the crossover from coherent lattice dynamics to incoherent many-body relaxation in a polar van der Waals semiconductor. The approach can be extended to other low-dimensional ferroic materials in which electronic states couple strongly to low-frequency structural coordinates. Combining ultrafast spectroscopy with temperature, carrier density, strain or electric-field control may enable selected phonons to guide transient structural and electronic states, rather than merely record their relaxation.

%


\section*{Methods}

\subsection{Transient-grating optical measurements with experimental conditions.} 

Transient-grating (TG) measurements were performed using a diffractive-optics-based, all-reflective two-dimensional spectrometer, as described in our previous reports\cite{Ref36, Ref37}, with an intrinsic phase stability of approximately $\lambda$/160. Broadband excitation pulses were generated using a home-built noncollinear optical parametric amplifier (NOPA) pumped by a commercial femtosecond laser system (Spectra-Physics, Newport). The pulses were compressed to approximately 5 fs using a 19-channel deformable mirror (OKO Technologies) in combination with two pairs of chirped mirrors (Munich UltraFast Innovations GmbH). The temporal pulse profile was characterized by frequency-resolved optical gating (FROG), with retrieval performed using the FROG3 software package (Femtosecond Technologies). The resulting excitation spectrum had a full width at half maximum of $\sim$200 nm, centred at 620 nm, providing sufficient spectral bandwidth to access the relevant electronic transitions of $\alpha$-In$_{2}$Se$_{3}$. For the nonlinear optical measurements, three excitation beams were focused onto the sample to a spot diameter of approximately 130 {\rm $\mu$}m, generating the transient-grating signal in the appropriate phase-matching direction. The emitted signal was dispersed using a Sciencetech 9055F spectrometer and detected with a CCD linear-array detector (Entwicklungsbüro Stresing). Unless otherwise stated, measurements were performed at a repetition rate of 1 kHz with the excitation pulse energy attenuated to approximately 10 nJ at the sample. The phase of the measured TG response was determined using the invariant-theorem procedure described previously in Ref.\ \cite{Ref38}.

\subsection{First-principles calculations}

First-principles calculations were performed within density-functional theory\cite{Ref39,Ref40} using the projector augmented-wave method\cite{Ref41,Ref42} and the Perdew--Burke--Ernzerh\cite{Ref43} of exchange--correlation functional, as implemented in VASP\cite{Ref44}. Bulk 3R-$\alpha$-In$_2$Se$_3$ was represented by a 15-atom hexagonal cell. The lattice parameters and atomic positions were relaxed using a plane-wave cutoff of 500~eV and a $12\times12\times6$ Monkhorst--Pack mesh. Grimme's D3 dispersion correction with Becke--Johnson damping was included during structural relaxation. The electronic convergence threshold was $10^{-8}$~eV, and the residual-force threshold was 0.01~eV~\AA$^{-1}$. Zone-centre phonon frequencies and eigenvectors were obtained using central finite differences with atomic displacements of 0.01~\AA, a cutoff of 300~eV, and a $\Gamma$-centred $12\times12\times6$ mesh. Non-resonant Raman activities were evaluated from finite differences of the macroscopic dielectric tensor along the normal-mode coordinates. Electronic band dispersions were calculated with a cutoff of 500~eV along the reciprocal-space path specified in SI Sec.~\Rmnum{5}.

%
\begin{addendum}

\item[Acknowledgements]
This work was supported by the National Key Research and Development Program of China grant no.~2024YFA1409800)(H.-G.D. and X.D.), NSFC Grant with no.~12274247 (H.-G.D), no.~W2421021 (H.-G.D.) and no.~12504279 (F.Z.), the Zhejiang Provincial natural Science Foundation of China with nos.~LQn26A040009 (F.Z.) and LBMhQn26e020006 (J.Z.), the Yongjiang talents program with no.~2022A-094-G (H.-G.D.) and no.~2023A-158-G (F.Z.), Ningbo international Science and Technology Cooperation with no.~2023H009 (H.-G.D.), and ``lixue+'' innovation Leading Project and the foundation of national excellent young scientist (H.-G.D.). the Next Generation Chemistry theme at the Rosalind Franklin institute is supported by the EPSRC [V011359/1 (P)] (A.J.).

\item[Supplementary information]
Supplementary Information accompanies this manuscript and provides additional details of sample preparation, Raman-spectrum comparisons, phonon eigenvectors, electronic-structure calculations, hierarchical equations of motion, wavelet analysis, and the effective vibronic model.

\item[Competing interests]
The authors declare no competing interests.

\item[Correspondence]
Correspondence of paper should be addressed to H.-G.D. (email: duanhongguang@nbu.edu.cn) and A.J. (Ajay.Jha@rfi.ac.uk).

\end{addendum}
%
\newpage

\newpage

\newpage

\newpage

\end{document}


\title{Supplementary Information \\[3mm] Long-lived Coherent Phonons Reveal Competing Thermal and Many-Body Dynamics in $\alpha$-In$_{2}$Se$_{3}$}  

\author{Xuanchao Zhang$^{1,*}$, Dehao Yuan$^{1,*}$, Junhua Zhou$^{1}$, Vandana Tiwari$^{2}$, Fulu Zheng$^{1}$, Ajay Jha$^{3,4}$, Hong-Guang Duan$^{1}$} 

\affiliation{
$^{1}$Department of Physics, School of Physical Science and Technology, Ningbo University, Ningbo 315211, China \\ 
$^{2}$Linac Coherent Light Source, SLAC National Accelerator Laboratory, Menlo Park, California 94025 \\ 
$^{3}$Rosalind Franklin Institute, Harwell, Oxfordshire OX11 0QX, United Kingdom \\  
$^{4}$Department of Pharmacology, University of Oxford, Oxford, OX1 3QT United Kingdom \\ 
} 
\date{\today} 

\maketitle 

This Supplementary Information provides details of sample preparation, calculated Raman spectra and phonon eigenvectors, the HEOM formalism, electronic-structure calculations, continuous-wavelet analysis, and the effective three-level vibronic model used to simulate the transient-grating response.

\section{sample preparation}

The 3R structure $\alpha$ phase In$_2$Se$_3$ single crystal was ordered from 2D semiconductor, the website: https://2dsemiconductors.com/. We used it without further modification. The thin slide of sample has been prepared by microtome machine (Leica EM UC7 Ultramicrotome) and cut to 50 nm thickness, which allows the optical beam passing through the sample with transmission configuration of measurements. Moreover, the thin sample was further mounted on a glass substrate (Quartz) with thickness of 1 mm. The sample with substrate was further mounted in a H2 closed cycle cryostat system (US ARS cryostat), which allows us to perform the temperature-dependent measurements. The sample and cryostat has been further installed in a XYZ system, which allows to adjust the sample and laser spot positions and reduce the scattering during the measurement. The optical measurement has been written in the next subsection. 

\section{Calculated Raman Spectrum and Comparison with Experiment}

\begin{figure}[H]
\centering
\includegraphics[width=12cm]{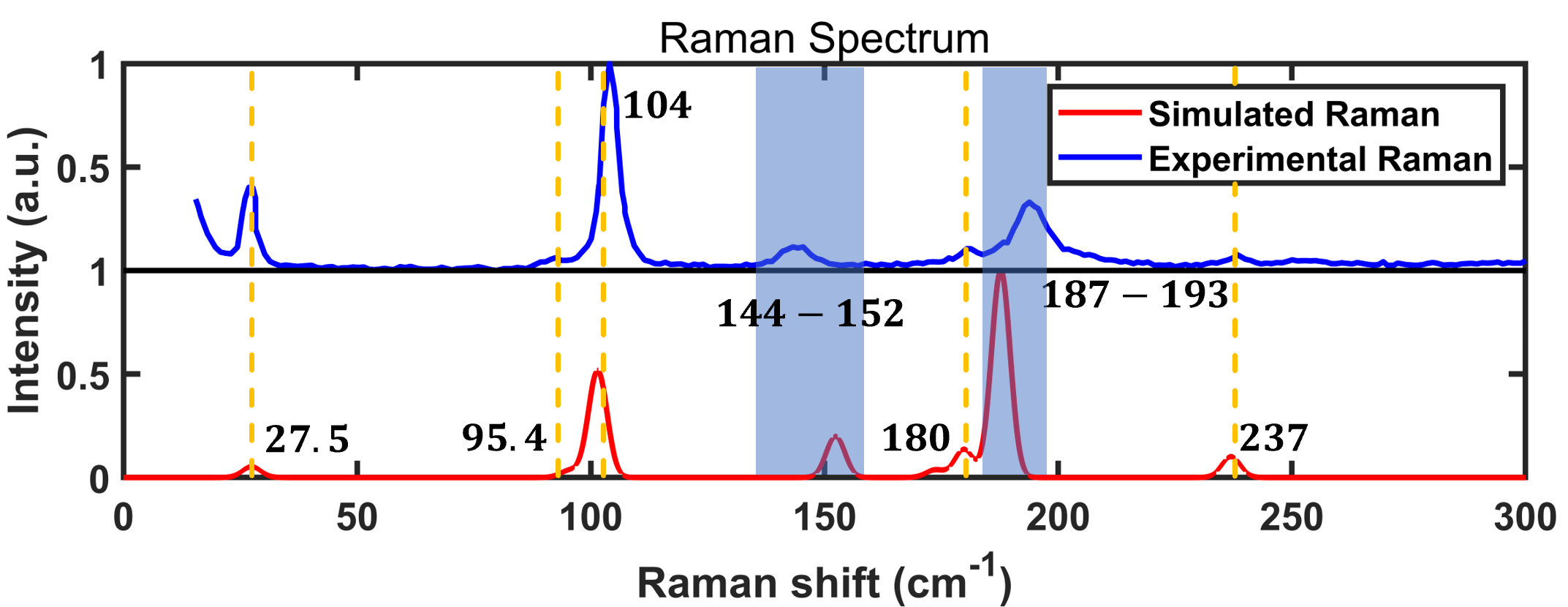} 
\caption{Comparison of the calculated and experimental Raman spectra of $\alpha$-In$_{2}$Se$_{3}$. The simulated Raman spectrum (red) is compared with the experimental spectrum (blue) \cite{Raman}, with both spectra normalized to their respective maximum intensities and vertically offset for clarity. The orange dashed lines indicate the calculated Raman-active mode frequencies, while the shaded region highlights the spectral range around 135–160 cm$^{-1}$.}
\label{fig:FigS1}
\end{figure}

The calculated Raman spectrum of $\alpha$-In$_{2}$Se$_{3}$ is shown in Fig.\ \ref{fig:FigS1}. Several Raman-active vibrational modes are obtained below 250 cm$^{-1}$, with prominent calculated peaks located at approximately 27.5, 101.47, 152.4, 179.3, 187.6, and 236.8 cm$^{-1}$. Among these modes, the strongest calculated Raman response occurs near 187.6 cm$^{-1}$, while additional pronounced features are observed near 101.47 and 152.4 cm$^{-1}$. The calculated spectrum is compared with the polarized Raman experiment. Experimentally, Raman features were observed at approximately 27, 91, 104, 144, 180-182, 187, 193-203, and 237 cm$^{-1}$, depending on the scattering geometry and polarization configuration. the calculated Raman frequencies show reasonable agreement with the experimental spectrum. In particular, the calculated modes at 27.5, 179.3, and 236.8 cm$^{-1}$ closely reproduce the experimental features near 27, 180-182, and 237 cm$^{-1}$, respectively. The calculated feature at 152.4 cm$^{-1}$ is located at a somewhat higher frequency than the experimental feature near 144 cm$^{-1}$. Additional calculated modes around 101.47 and 187.6 cm$^{-1}$ fall within frequency regions where multiple experimental Raman features are observed.

Despite these differences in measurement and calculation conditions, the overall calculated frequency distribution is consistent with the experimental Raman data. Both the theoretical and experimental spectra show that the dominant first-order Raman-active vibrational modes of $\alpha$-In$_{2}$Se$_{3}$ are concentrated below approximately 250 cm$^{-1}$. In particular, the agreement near 27, 180, and 237 cm$^{-1}$ provides direct correspondence between several calculated and experimentally observed modes. The remaining frequency differences are moderate and are consistent with the expected limitations of harmonic first-principles phonon calculations and the differences between an idealized theoretical crystal and a room-temperature experimental sample.

Overall, the comparison supports the validity of the calculated vibrational spectrum as a representation of the Raman-active phonon structure of $\alpha$-In$_{2}$Se$_{3}$. Further assignment of individual modes can be performed by analyzing the calculated mode symmetries and atomic displacement patterns and comparing them with the experimentally reported Raman modes.

\section{phonon mode of $\alpha$-In$_{2}$Se$_{3}$}

\begin{figure}[H]
\centering
\includegraphics[width=12cm]{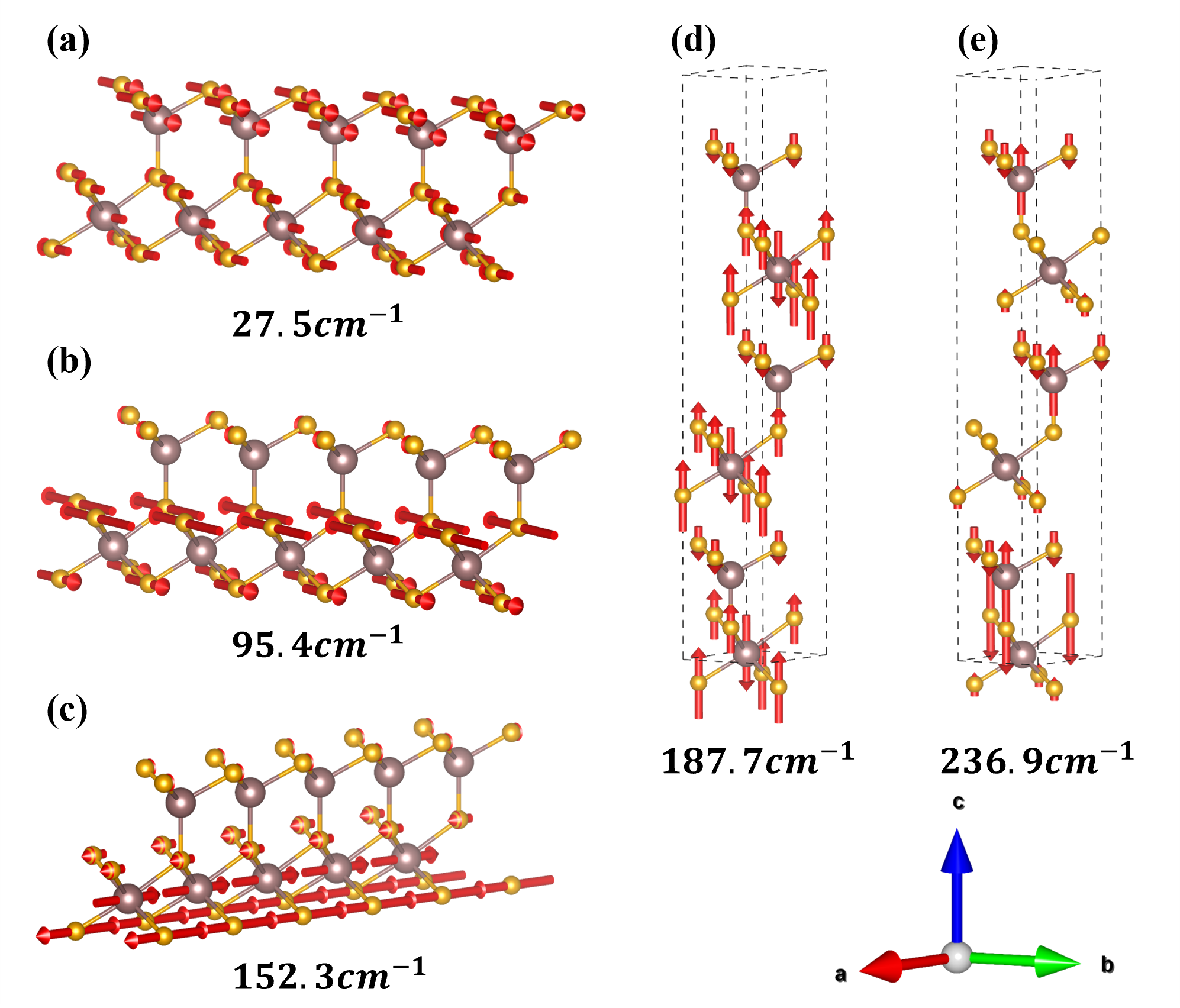} 
\caption{Calculated atomic displacement patterns of selected Raman-active vibrational modes of $\alpha$-In$_{2}$Se$_{3}$ at (a) 27.5 cm$^{-1}$, (b) 95.4 cm$^{-1}$, (c) 152.3 cm$^{-1}$, (d) 187.7 cm$^{-1}$ and (e) 236.9 cm$^{-1}$. Red arrows indicate the directions of atomic displacements associated with each normal mode. The crystallographic $a$, $b$ and $c$ directions are indicated at the bottom right.}
\label{fig:FigS2}
\end{figure}

The atomic displacement patterns of other phonon modes of $\alpha$-In$_{2}$Se$_{3}$ are shown in Fig.\ \ref{fig:FigS2} representative modes with calculated frequencies of 27.5, 95.4, 152.3, 187.7, and 236.9 cm$^{-1}$ were selected to illustrate the characteristic lattice vibrations over the frequency range of the calculated Raman spectrum. The red arrows indicate the direction and relative displacement of the atoms associated with each normal mode.

The calculated modes exhibit distinct atomic displacement patterns as the vibrational frequency increases. The low-frequency modes at 27.5, 95.4 and 152.3 cm$^{-1}$ involve relatively collective atomic motions extending over the layered structure. The mode at 101.47~cm$^{-1}$, shown in Fig.~4b of the main text, is dominated by atomic displacements along the layer-normal direction. By contrast, the mode labelled 95.4~cm$^{-1}$ in Fig.~\ref{fig:FigS2}(b) occur predominantly in-plane atomic displacements. Thus, the 95.4 and 101.47~cm$^{-1}$ labels refer to distinct normal modes. The proposed assignment of the experimentally dominant oscillation near 104~cm$^{-1}$ concerns the 101.47~cm$^{-1}$ mode. At higher frequencies, the modes at 187.7 and 236.9 cm$^{-1}$ are characterized by more pronounced relative displacements of neighboring atoms, with substantial displacement components along the crystallographic $c$ direction. These mode-resolved displacement patterns provide additional information for interpreting the individual features observed in the calculated Raman spectrum and for comparison with experimentally reported Raman modes of $\alpha$-In$_{2}$Se$_{3}$.

\section{Hierarchy equation of motion} 

To accurately describe nonadiabatic dynamics in the presence of environmental dissipation, we employ the Hierarchy Equation of Motion (HEOM) formalism. HEOM provides a numerically exact treatment of system-bath interactions by explicitly incorporating bath-induced memory effects, enabling reliable simulations of population and coherence dynamics beyond perturbative or Markovian approximations. In this section, we show detailed information on
HEOM. For this, the well-known bath correlation function is given as 
\begin{equation}
\begin{split}
C(t)=\sum_{k=0}^{\infty}\left(c_ke^{-v_kt}\right),	
\end{split}
\end{equation} 
where $v_{0}=\gamma,v_{k>0}=\frac{2\pi k}{\beta\hbar}$ is Matsubara frequency. The parameter $c_{k}$ is
\begin{equation}
\begin{split}
c_0=\gamma\lambda[\cot(\frac{\beta\hbar\gamma}{2})-i],
\end{split}
\end{equation} 
\begin{equation}
\begin{split}
c_{k>0}=\frac{4\gamma^2}{\beta\hbar}\frac{v_k}{v_k^2-\gamma^2}.
\end{split}
\end{equation} 
Under the separable initial conditions of the system and bath, the path integral form of the reduced density matrix of the three-level system is 
\begin{equation}
\begin{split}
\rho(\bar{\psi},\psi^{\prime};t)=\int D\bar{\psi}D\psi\int D\bar{\psi}^{\prime}D\psi^{\prime}\rho(\bar{\psi}_{0},\psi_{0}^{\prime};t_{0})e^{\frac{iS[\bar{\psi},\psi]}{\hbar}}\mathcal{F}[\bar{\psi},\psi;\bar{\psi}^{\prime},\psi^{\prime}]e^{\frac{-iS[\bar{\psi},\psi]}{\hbar}},
\end{split}
\end{equation} 
where, $\psi$ and $\bar{\psi}$ are Grassmann algebra, and coherent state $\left|\psi\right\rangle$ is satisfied 
\begin{equation}
\begin{split}
\hat{\psi}\left|\psi\right\rangle=\psi\left|\psi\right\rangle,
\end{split}
\end{equation} 
\begin{equation}
\begin{split}
\langle\psi|\hat{\psi}^\dagger=\bar{\psi}\langle\psi|.
\end{split}
\end{equation} 
Here, $\mathcal{F}[\bar{\psi},\psi;\bar{\psi}^{\prime},\psi^{\prime}]$ is Feynman-Vernon Influence Function, its expression is 
\begin{equation}
\begin{split}
\mathcal{F}[\bar{\psi},\psi;\bar{\psi}^{\prime},\psi^{\prime}] 
=\exp(-\frac{1}{\hbar}\int_{0}^{\infty}d\omega J(\omega)\int_{t_{0}}^{t}ds\int_{t_{0}}^{s}ds^{\prime}V^{\times}(s)[V^{\times}(s^{\prime})\coth(\frac{\beta\hbar\omega}{2})\cos(\omega(s-s^{\prime}))-iV^{\circ}(s)\sin(\omega(s-s^{\prime}))]).
\end{split}
\end{equation} 
Abbreviations are used 
\begin{equation}
\begin{split}
V^{\times}(t)\equiv V(\bar{\psi}(t),\psi(t))-V(\bar{\psi}^{\prime}(t),\psi|^{\prime}(t)),
\end{split}
\end{equation} 

\begin{equation}
\begin{split}
V^{\circ}(t)\equiv V(\bar{\psi}(t),\psi(t))+V(\bar{\psi}^{\prime}(t),\psi^{\prime}(t)).
\end{split}
\end{equation} 
In order to derive the equation of motion, the auxiliary density matrix is introduced as 
\begin{equation}
\begin{split}
& \rho_{j_{1},\ldots,j_{k}}^{(n)}(\bar{\psi},\psi^{\prime};t)=\int D\bar{\psi}D\psi\int D\bar{\psi}^{\prime}D\psi^{\prime}\rho(\bar{\psi}_{0},\psi_{0}^{\prime};t_{0}) \\
& \times\{e^{-\gamma t}[-\int_{t_0}^tds\gamma\Theta(s)e^{\gamma s}]\}^n \\
& \times\prod_{k=1}^{K}\{e^{-v_{k}t}[-\int_{t_{0}}^{t}dsv_{k}\psi_{k}(s)e^{v_{k}s}]\}^{j_{k}} \\
& \times e^{\frac{iS[\bar{\psi},\psi]}{\hbar}}\mathcal{F}[\bar{\psi},\psi;\bar{\psi}^{\prime},\psi^{\prime}]e^{\frac{-iS[\bar{\psi},\psi]}{\hbar}},
\end{split}
\end{equation} 
where, $n, j_1,\ldots, j_k$ are nonnegative integer. By expanding the auxiliary density matrix by derivation, the desired hierarchy equation of motion (HEOM) can be written as 
\begin{equation}
\begin{split}
\label{eq:s100} 
\frac{\partial}{\partial t}\rho_{j_{1},...,j_{K}}^{(n)}(t)= & -[i\hat{\mathcal{L}}+n\gamma+\sum_{k=1}^{K}(j_{k}v_{k}+\hat{\phi}\hat{\psi}_{k})+\Xi]\rho_{j_{1},...,j_{k}}^{(n)}(t) \\
& -\hat{\phi}\rho_{j_{1},...,j_{K}}^{(n+1)}(t) \\
& -n\gamma\hat{\Theta}\rho_{j_{1},\ldots,j_{K}}^{(n-1)}(t) \\
& -\sum_{k=1}^{K}\hat{\phi}\rho_{j_{1},...,j_{k}+1,...,j_{K}}^{(n)}(t) \\
& -\sum_{k=1}^{K}j_{k}v_{k}\hat{\psi}_{k}\rho_{j_{1},...,j_{k}-1,...,j_{K}}^{(n)}(t),
\end{split}
\end{equation} 
where, $\mathcal{L}$ is the Liouvillian of the three-level system. The relaxation operator $\hat{\phi},\hat{\Theta}\mathrm{~and~}\hat{\psi}_k\text{ are obtained by }V^\times\mathrm{~and~}V^\circ.$  

The expression are 
\begin{equation}
	\begin{split}
 \phi\equiv iV^{\times},
	\end{split}
\end{equation}
\begin{equation}
	\begin{split}
 \Theta\equiv\frac{i\zeta}{\beta\hbar\omega_{0}}\left[-i\frac{\beta\hbar\gamma}{2}V^{\circ}+\frac{\beta\hbar\gamma}{2}\cot\left(\frac{\beta\hbar\gamma}{2}\right)V^{\times}\right],
	\end{split}
\end{equation}
\begin{equation}
	\begin{split}
 \psi_{k}\equiv\frac{i\zeta}{\beta\hbar\omega_{0}}\frac{2\gamma^{2}}{v_{k}^{2}-\gamma^{2}}V^{\times} ,
	\end{split}
\end{equation}
\begin{equation}
	\begin{split}
v_{k}=\frac{2\pi k}{\beta\hbar},
	\end{split}
\end{equation}
\begin{equation}
	\begin{split}
 \hat{\Xi}\equiv\frac{\zeta}{\beta\hbar\omega_{0}}\left[1-\frac{\beta\hbar\gamma}{2}\cot\left(\frac{\beta\hbar\gamma}{2}\right)\right]\hat{V}^{\times}\hat{V}^{\times},
	\end{split}
\end{equation}
In the above equation, two hyper-operators are introduced. For any operator $\hat{\mathcal{O}}$ and operand $\hat{f}$ are satisfied
\begin{equation}
	\begin{split}
		\hat{\mathcal{O}^{\times}}\hat{f}\equiv\hat{\mathcal{O}}\hat{f}-\hat{f}\hat{\mathcal{O}^{\dagger}},
	\end{split}
\end{equation}
\begin{equation}
	\begin{split}
		\hat{\mathcal{O}}^{\circ}\hat{f}\equiv\hat{\mathcal{O}}\hat{f}+\hat{f}\hat{\mathcal{O}}^{\dagger}.
	\end{split}
\end{equation}
HEOM involves an infinite number of terms representing higher-order non-Markovian effects. In practice, the hierarchy
is truncated when the contribution of these terms becomes negligible. This ensures computational efficiency while
maintaining accuracy in capturing the system’s dynamics. To truncate Eq. \ref{eq:s100}, we introduce the terminator
\begin{equation}
	\begin{split}
	\frac{\partial}{\partial t}\rho_{j_{1},...,j_{K}}^{(n)}(t)\simeq-(i\hat{\mathcal{L}}+\sum_{k=1}^{K}\hat{\phi}\hat{\psi}_{k}+\hat{\Xi})\rho_{j_{1},...,j_{K}}^{(n)}(t),
	\end{split}
\end{equation}
and its truncation condition is
\begin{equation}
	\begin{split}
N\equiv n+\sum_{k=1}^{K}j_{k}\gg\frac{\omega_{0}}{\min(\gamma,v_{1})}.
	\end{split}
\end{equation} 



\section{Electronic band structure of $\alpha$--$In_{2}Se_{3}$} 

The electronic bands of bulk 3R-$\alpha$-In$_{2}$Se$_{3}$ were calculated using the projector augmented-wave method and the PBE generalized-gradient approximation, as implemented in VASP. The relaxed 15-atom hexagonal representation of the 3R unit cell and a plane-wave cutoff energy of 500~eV were used. Starting from the converged charge density, a spin-unpolarized line-mode non-self-consistent calculation was performed along $\Gamma$--F--K--$\Gamma$, with 101 calculated $k$ points per segment. Here, $\Gamma=(0,0,0)$, $\mathrm{F}=(1/2,0,0)$, and $\mathrm{K}=(1/3,1/3,0)$ in the reciprocal basis of the hexagonal cell. The energies are referenced to the valence-band maximum along the plotted path.

Along this $k_z=0$ path, the valence-band maximum occurs on the $\Gamma$--F segment near $\mathbf{k}=(0.185,0,0)$, whereas the conduction-band minimum occurs at $\Gamma$. The resulting indirect PBE Kohn--Sham gap is 0.500~eV, while the direct gap at $\Gamma$ is 0.644~eV. A separate calculation on a uniform $12\times12\times6$ mesh gives a similar coarse indirect gap of 0.501~eV but places the valence-band maximum slightly outside the $k_z=0$ plane. The plotted path therefore describes the in-plane band dispersion but does not constitute a complete search for band extrema throughout the three-dimensional Brillouin zone.

Spin--orbit coupling, quasiparticle corrections, optical transition matrix elements, and electron--hole interactions were not included. Consequently, the calculated Kohn--Sham gap should not be interpreted as the experimental optical gap, and the band dispersion alone does not identify the transitions dominating the broadband transient-grating response. The three-level vibronic model introduced in Sec.~\Rmnum{4} is therefore treated as an effective phenomenological representation rather than a one-to-one mapping of individual Kohn--Sham bands or $k$ points.

\begin{figure}[H]
\centering
\includegraphics[width=12cm]{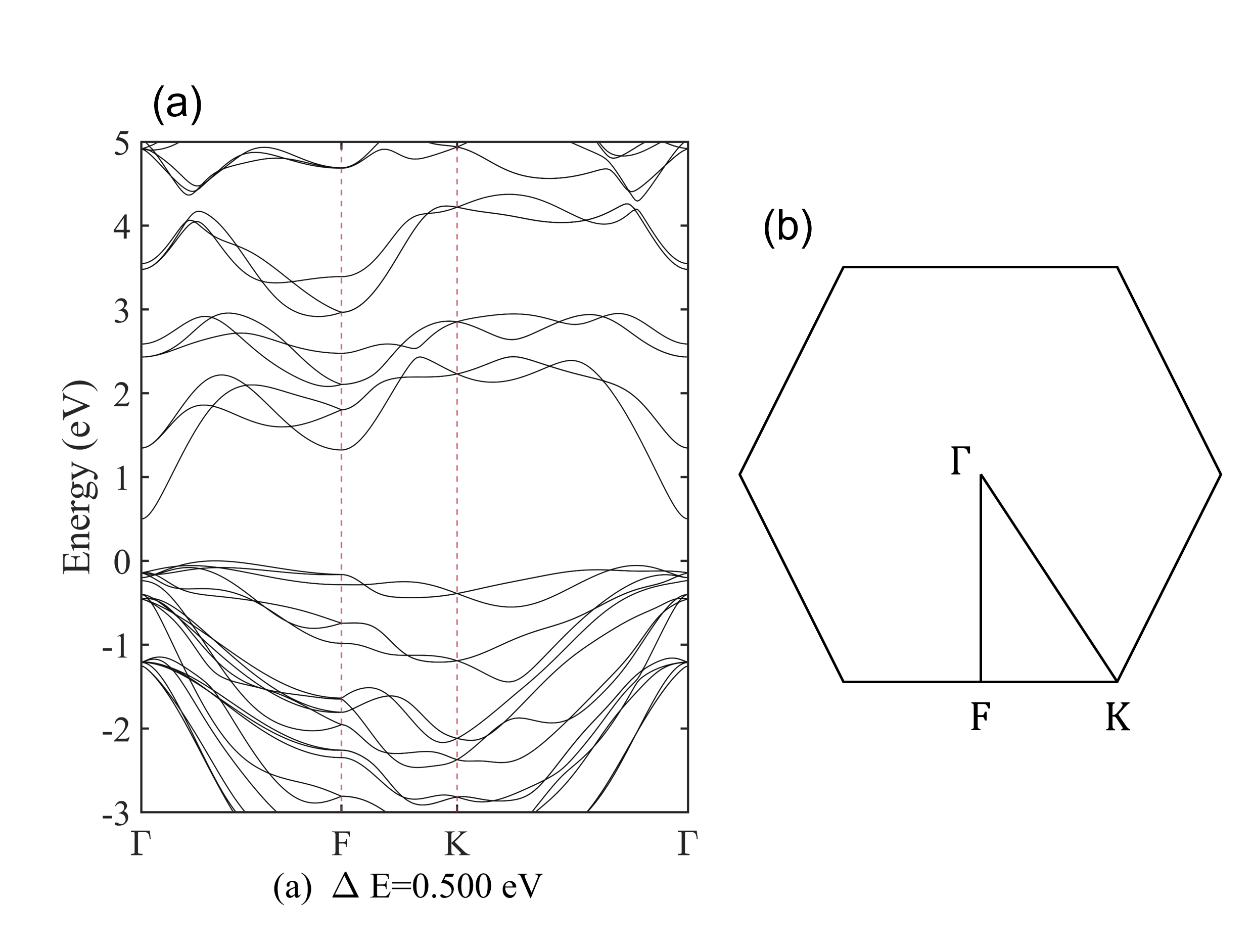} 
\caption{Calculated electronic band structure and reciprocal-space path of bulk 3R-$\alpha$-In$_{2}$Se$_{3}$.(a) Spin-unpolarized PAW-PBE band structure calculated without spin--orbit coupling along $\Gamma$--F--K--$\Gamma$. The energy zero is the valence-band maximum along the plotted path. The annotated $\Delta E=0.500$~eV is the indirect Kohn--Sham gap along this path, rather than an optical gap; the red dashed lines mark F and K.(b) Schematic of the $k_z=0$ reciprocal-space section showing the same path, with $\mathrm{F}=(1/2,0,0)$ and $\mathrm{K}=(1/3,1/3,0)$ in the reciprocal basis used for the calculation.}
\label{fig:FigS3}
\end{figure}

\section{ WAVELET ANALYSIS } 
In this section, we summarize the basic principles of the wavelet transform, the details of which can be found in Refs.\ \cite{wavelet1,wavelet2}. It starts with the definition of a zero-mean and short-time oscillating function $\psi$, called a "mother" wavelet, which is used to decompose a one-diemsional or multi-dimensional real-valued signals into different frequency bands. This mother wavelet function is translated in time by $t$ and stretched by the scaling factor $\omega^{-1}$ to obtain the "atomic" functions of the wavelet

\begin{eqnarray}
	\begin{aligned}
		psi_{t,\omega}(t')=\sqrt{\omega}\psi([t'-t]\omega).
	\end{aligned}
\end{eqnarray}

It provides an efficient basis for the transformation. The two most common transforms are the discrete wavelet transform and the continuous wavelet transform $CWT$ \cite{wavelet3}. Discrete transform divied a signal into several frequency bands and are often used for data and image compression. The continuous transform used in this paper is based on the expansion of the time signal $f(t)$ and is realized by the dot product of a function and a wavelet atom,

\begin{eqnarray}
	\begin{aligned}
		CWT_{f}(t,\omega)=\int^{+\infty}_{-\infty}dt'f(t')\sqrt{\omega}\psi^{*}([t'-t]\omega).
	\end{aligned}
\end{eqnarray}

The parameter $t$ represents the center position of the wavelet atom and the scaling parameter $\omega^{-1}$ controls the width of the wavelet atom relative to the mother wavelet function. This nonlinear integral transform provides high time resolution for high-frequency components, but high frequency resolution for the slowly changing signal components. Project the signal onto basis functions with differe ranges fixed at different "center" frequencies and scale fators.

\section{three-level vibronic model and transient-grating reponse}

To describe the electronic--vibrational dynamics underlying the calculated transient-grating (TG) response, we employed an effective three-level model comprising the electronic ground state $\ket{g}$, a singly excited state $\ket{e_1}$, and a higher excited state $\ket{e_2}$. The latter accounts for the excited state absorption part. This reduced model is phenomenological and is not intended as a one-to-one assignment of the effective electronic states to individual Kohn--Sham bands or crystal momenta.

The molecular Hamiltonian is
\begin{equation}
\begin{split}
H_{\mathrm{mol}}={}&\ket{g}h_g\bra{g}
+\ket{e_1}h_{e_1}\bra{e_1}
+\ket{e_2}h_{e_2}\bra{e_2}\\
&+V_0\left(\ket{g}\bra{e_1}+\ket{e_1}\bra{e_2}+\mathrm{H.c.}\right),
\end{split}
\label{eq:vibronic_hamiltonian}
\end{equation}
where
\begin{equation}
h_g=\sum_{i=a,b}\frac{\Omega_i}{2}\left(P_i^2+Q_i^2\right),
\end{equation}
and
\begin{equation}
h_{e_j}=\epsilon_j+h_g+\sqrt{2}\,\kappa_{a,j}Q_a
+\sqrt{2}\,\kappa_{b,j}Q_b,
\qquad j=1,2.
\label{eq:excited_pes}
\end{equation}
Here, $Q_i$ and $P_i$ are dimensionless vibrational coordinates and conjugate momenta. All Hamiltonian parameters are expressed in cm$^{-1}$. For the baseline calculation used to generate the response in Fig.~4 of the main text, the two effective modes were assigned frequencies $\Omega_a=101.47$~cm$^{-1}$ and $\Omega_b=187.7$~cm$^{-1}$. The electronic energies were $\epsilon_1=15800$~cm$^{-1}$ and $\epsilon_2=33721$~cm$^{-1}$, and the linear vibronic-coupling parameters were $(\kappa_{a,1},\kappa_{b,1})=(33.2,12.2)$~cm$^{-1}$ and $(\kappa_{a,2},\kappa_{b,2})=(62.4,23.7)$~cm$^{-1}$. A constant intermanifold coupling $V_0=100$~cm$^{-1}$ was used between adjacent electronic manifolds. The corresponding unperturbed transition energies, $\epsilon_1=15800$~cm$^{-1}$ and $\epsilon_2-\epsilon_1=17921$~cm$^{-1}$, lie near 633 and 558~nm, respectively. The nominal 101.47~cm$^{-1}$ coordinate represents the experimentally dominant oscillation near 104~cm$^{-1}$.

Within the Condon approximation, the lowering and raising components of the transition-dipole operator are
\begin{equation}
\mu^-=\left(
\mu_{g1}\ket{g}\bra{e_1}
+\mu_{12}\ket{e_1}\bra{e_2}
\right)\otimes I_{\mathrm{vib}},
\qquad
\mu^+=(\mu^-)^{\dagger},
\end{equation}
with relative transition-dipole amplitudes $\mu_{g1}=1$ and $\mu_{12}=\sqrt{2}$. No direct $g\leftrightarrow e_2$ transition was included. Optical pathways involving the $g\leftrightarrow e_1$ transition contribute to ground-state bleaching and stimulated emission, whereas pathways involving the $e_1\leftrightarrow e_2$ transition give the excited-state-absorption contribution.

\begin{figure}[H]
\centering
\includegraphics[width=14cm]{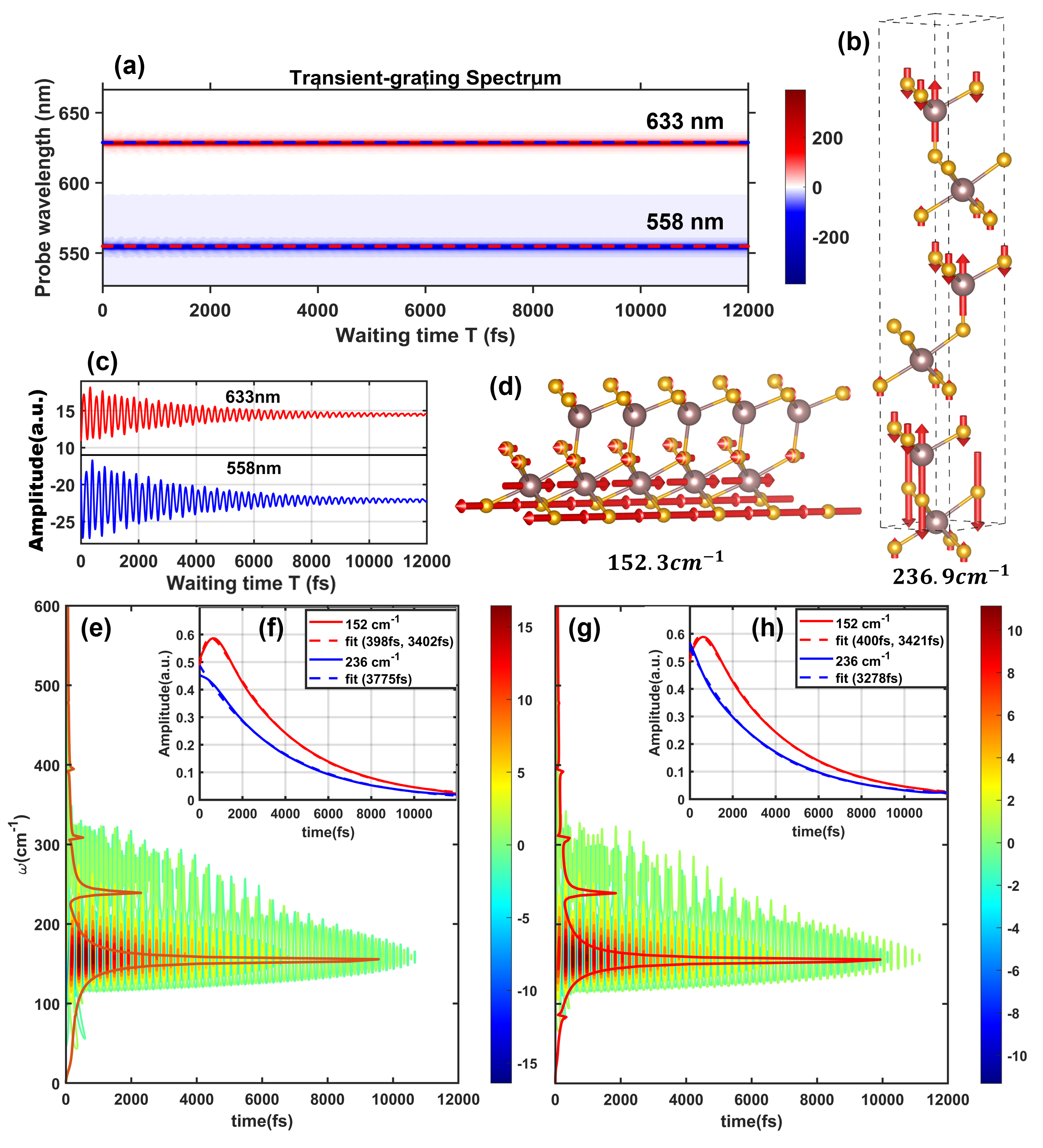}
\caption{Wavelet analysis of the calculated transient-grating response and mode-substitution test. (a) Calculated TG spectrum for the alternative parameterization using modes near 152 and 236~cm$^{-1}$. (b,d) Calculated $\Gamma$-point phonon eigenvectors of $\alpha$-In$_2$Se$_3$ at 236.9 and 152.3 cm$^{-1}$. (c) Population dynamic maps of the mode-substituted TG waiting-time traces at 558~nm and 633~nm, showing components near 152 and 237~cm$^{-1}$. (e,g) Corresponding wavelet maps at (e) 558~nm and (g) 633~nm. (f,h) Extracted amplitudes and fits of the 152~cm$^{-1}$ (red) and 236~cm$^{-1}$ (blue) components.}
\label{fig:FigS4}
\end{figure}

The continuous-wavelet procedure described in Sec.~\Rmnum{6} was applied to the calculated TG waiting-time traces. For the baseline parameterization, the wavelet maps and amplitude profiles are now presented in Fig.~4(f--i) of the main text rather than in Fig.~\ref{fig:FigS4}. The nominal Hamiltonian frequencies are 101.47 and 187.7~cm$^{-1}$, whereas the maxima extracted from the finite-duration traces occur near 101 and 179~cm$^{-1}$, respectively. The low-frequency profiles were described empirically by $A_{\mathrm{bi}}(T)=A_1\exp(-T/\tau_1)+A_2\exp(-T/\tau_2)+C$, and the high-frequency profiles by $A_{\mathrm{mono}}(T)=A_0\exp(-T/\tau)+C$. At 558~nm, the baseline fits give $(\tau_1,\tau_2)=(495,2295)$~fs for the $\sim101$~cm$^{-1}$ component and $\tau=1742$~fs for the $\sim179$~cm$^{-1}$ component. At 633~nm, the corresponding values are $(485,2313)$~fs and 1954~fs. These constants provide empirical measures of the simulated wavelet envelopes rather than unique microscopic phonon lifetimes.

To assess how the calculated response depends on the selected vibrational coordinates, we performed a mode-substitution test using $\Omega_a=152.3$~cm$^{-1}$ and $\Omega_b=236.7$~cm$^{-1}$, representing the calculated modes labelled 152.3 and 236.9~cm$^{-1}$ in Fig.~\ref{fig:FigS4}. The vibronic couplings were $(\kappa_{a,1},\kappa_{b,1})=(20.26,17.86)$~cm$^{-1}$ and $(\kappa_{a,2},\kappa_{b,2})=(38.28,33.74)$~cm$^{-1}$, while the electronic energies, intermanifold coupling $V_0$, and transition-dipole ratios were retained from the baseline calculation. Consequently, the alternative TG spectrum retains the oppositely signed bands near 633 and 558~nm [Fig.~\ref{fig:FigS4}(a)], and the associated waiting-time traces are shown in Fig.~\ref{fig:FigS4}(c). The displacement patterns in Fig.~\ref{fig:FigS4}(b,d) correspond to the 236.9 and 152.3~cm$^{-1}$ modes, respectively; the former contains substantial layer-normal motion, whereas the latter has a stronger in-plane component.

Environmental dissipation was described using two bath channels with identical bath parameters. In the wavenumber representation, the Drude--Lorentz spectral density is written as
\begin{equation}
J_\alpha(\nu)= \frac{2\lambda_\alpha\gamma_\alpha\nu}
{\nu^2+\gamma_\alpha^2},
\qquad \alpha=1,2,
\end{equation}
where $\gamma_\alpha$ is the bath cutoff and $\lambda_\alpha$ is the reorganization parameter for the adopted system--bath coupling operator normalization. The implementation uses the equivalent
parameterization $\lambda_\alpha=\zeta_\alpha\gamma_\alpha/(2\Omega_0)$. With $\zeta_\alpha=0.2$ and $\Omega_0=1$ in the simulation normalization, $\zeta_\alpha/\Omega_0=0.2$ and $\lambda_\alpha=5$~cm$^{-1}$.

The bath correlation functions were treated within the high-temperature approximation, retaining one Drude exponential per bath without explicit Matsubara terms or a Matsubara-tail correction. The inverse thermal energy in wavenumber units, $\widetilde{\beta}=hc/(k_{\mathrm B}T)$, was set to 0.0047959~cm, corresponding to $T\simeq300$~K. The bath parameters are summarized in Table~\ref{tab:HEOM_parameters}.

\begin{table}[htbp]
\centering
\caption{Bath parameters used in the high-temperature HEOM calculations. The normalized coupling
$\eta_\alpha=\zeta_\alpha/\Omega_0$ corresponds to the code inputs $\zeta_\alpha=0.2$ and $\Omega_0=1$. The equivalent Drude--Lorentz reorganization parameter is $\lambda_ \alpha = \eta_ \alpha \gamma_ \alpha/2$.}
\label{tab:HEOM_parameters}
\small
\begin{tabular}{lcc}
\hline
Parameter & Bath 1 & Bath 2 \\
\hline
Temperature $T$ (K)
& 300 & 300 \\
Inverse thermal energy $\widetilde{\beta}$ (cm)
& 0.0047959 & 0.0047959 \\
Bath cutoff $\gamma_\alpha$ (cm$^{-1}$)
& 50.0 & 50.0 \\
Normalized coupling $\eta_\alpha=\zeta_\alpha/\Omega_0$
& 0.2 & 0.2 \\
Reorganization parameter $\lambda_\alpha$ (cm$^{-1}$)
& 5.0 & 5.0 \\
\hline
\end{tabular}
\end{table}

The continuous-wavelet maps and extracted amplitude profiles are shown for 558~nm in Fig.~\ref{fig:FigS4}(e,f) and for 633~nm in Fig.~\ref{fig:FigS4}(g,h). At 558~nm, the 152~cm$^{-1}$ component is described by a biexponential envelope with $(\tau_1,\tau_2)=(398,3402)$~fs, whereas the 236~cm$^{-1}$ component gives a monoexponential constant of $\tau=3775$~fs. At 633~nm, the corresponding values are $(\tau_1,\tau_2)=(400,3421)$~fs and $\tau=3278$~fs. Retaining the electronic parameters leaves the spectral positions of the 558 and 633~nm bands essentially unchanged, whereas replacing the vibrational pair shifts the wavelet ridges to approximately 152 and 236~cm$^{-1}$. The calculated oscillatory frequencies therefore track the vibrational coordinates explicitly included in the effective Hamiltonian. This result constitutes a model-sensitivity test and does not supersede the assignment of the experimentally dominant $\sim104$~cm$^{-1}$ response to the calculated 101.47~cm$^{-1}$ mode.

